\documentclass[showpacs, showkeys,twocolumn,preprintnumbers,floatfix,amsmath,amssymb]{revtex4}
\usepackage{booktabs}
\usepackage{natbib}
\usepackage{amsmath,amsthm}
\usepackage{float}
\usepackage{graphicx}
\usepackage{epstopdf}
\usepackage{caption}
\usepackage{subfigure}
\usepackage{tabularx}
\usepackage{makecell}
\usepackage{threeparttable}
\usepackage{dcolumn}
\usepackage{bm}
\usepackage{color,epsfig,multirow}
\usepackage{array}
\usepackage{hyperref}
\usepackage{algorithm}
\usepackage{algpseudocode}
\usepackage{threeparttable}

\makeatletter
\def\@fnsymbol#1{\ensuremath{\ifcase#1\or *\or \ddagger\or \mathsection\or \mathparagraph\or \|\or **\or \dagger\dagger \or \ddagger\ddagger \else\@ctrerr\fi}}
\makeatother

\DeclareMathOperator{\erf}{erf}
\DeclareMathOperator{\erfc}{erfc}

\theoremstyle{definition}

\theoremstyle{remark}

\allowdisplaybreaks[4]

\begin{document}
\preprint{Preprint}

\title{RBMD 2.0: Random batch molecular dynamics package for large-scale simulations on multi-GPU architectures}
\author{Qi Zhou$^{1}$}
\author{Yongfa Guo$^{2}$}
\author{Jincheng Zhong$^{3}$}
\author{Shou-Hang Bo$^{3,4}$}
\author{Teng Zhao$^{2}$}\thanks{zhaoteng@sog-aitech.com}
\author{Zhenli Xu$^{1,2}$}

\affiliation{$^1$\mbox{School of Mathematical Sciences, Shanghai Jiao Tong University, Shanghai 200240, China.}} 
\affiliation{$^2$\mbox{SOG AI-Technology Co. Ltd., Shanghai, China.}}
\affiliation{$^3$ Future Battery Research Center, Global Institute of Future Technology, Shanghai Jiao Tong University, Shanghai 200240, China.}
\affiliation{$^4$ State Key Laboratory of Synergistic Chem-Bio Synthesis, School of Chemistry and Chemial Engineering, Shanghai Jiao Tong University, Shanghai 200240, China.}

\date{\today}
\begin{abstract}
 Large-scale molecular dynamics simulations of particle systems on multi-GPU architectures are often constrained by the computational and communication costs of nonbonded force evaluation. We present RBMD 2.0, a major new release of the random batch molecular dynamics package designed for cross-node multi-GPU simulations of large-scale systems. It combines the improved random batch Ewald method with three-dimensional domain decomposition and ghost-particle communication to accelerate multi-GPU nonbonded force evaluation, while the DTK CUDA framework facilitates portability across heterogeneous accelerator architectures. Numerical experiments on multiple benchmark systems demonstrate both the accuracy and efficiency of simulations with RBMD 2.0. For simulations involving up to hundreds of millions of particles across multiple accelerator devices, one achieves speedups ranging from severalfold to approximately two orders of magnitude in nonbonded force evaluation while exhibiting over $97.5\%$ weak-scaling behavior. These results demonstrate the promising nature of RBMD 2.0 as a computational engine for future exascale molecular dynamics simulations.
\end{abstract}




\maketitle

\section{Introduction}
\label{sec::intro}

Molecular dynamics (MD) simulation is one of the most important computational tools for bridging microscopic structures and macroscopic behavior \cite{karplus2002molecular,dror2012biomolecular,allen2017computer,frenkel2002understanding,tuckerman2010statistical}. Over the past several decades, substantial advances have been made in force-field development, time-integration algorithms, and massively parallel computing on multicore CPUs, enabling the widespread application of MD simulations in fields ranging from biomolecular systems and soft matter to materials science \cite{thompson2022lammps,abraham2015gromacs,phillips2020scalable,eastman2017openmm,salomon2013routine}. However, many scientifically important problems intrinsically involve large-scale systems
\cite{mindemark2018beyond,hallinan2013polymer,meyer1998polymer,
borodin2006mechanism,diddens2013lithium,kavokine2021fluids}.
When the simulated system is too small, interactions with periodic images and
finite-size hydrodynamic effects can introduce systematic biases into physical
observables, including diffusion coefficients, viscosities, polymer-chain
relaxation dynamics, and ion-cluster statistics
\cite{yeh2004system,jamali2018finite,celebi2021finite,kremer1990dynamics,
dunweg1993molecular,webb2015systematic}.
There is therefore a growing demand for molecular dynamics engines capable of
simulating millions to hundreds of millions of charged particles while retaining
the statistical reliability required for physically meaningful interpretation.

The challenges associated with such large-scale simulations are multifaceted.
First, the long-ranged nature of Coulomb interactions makes the computational cost of large-scale molecular dynamics simulations prohibitively high. Although widely used mesh-based methods, such as the
particle-particle particle-mesh (PPPM) \cite{hockney1988computer}, particle-mesh Ewald \cite{Darden1993JCP}, and
smooth particle-mesh Ewald methods \cite{essmann1995smooth}, reduce the computational complexity
to $O(N\log N)$ through fast Fourier transforms (FFTs), they require substantial
inter-node communication \cite{arnold2013comparison,ayala2021scalability,pall2020heterogeneous}. Second, even in the absence of long-range
interactions, conventional approaches for evaluating short-range forces,
including linked-cell and Verlet-list algorithms, introduce
considerable memory overhead \cite{plimpton2012computational,thompson2022lammps,hess2008gromacs,pall2020heterogeneous}. The limited memory capacity available on
individual computing nodes or accelerators can therefore directly constrain
the maximum system size accessible to large-scale MD simulations. Moreover, ongoing changes in high-performance computing architectures have
further intensified these challenges. Owing to their massive thread-level
parallelism, GPU clusters have become indispensable
hardware resources for contemporary exascale simulations \cite{hagerty2023studying,johansson2025lammpskokkos,beck2024all}. A central challenge
is therefore to develop fast algorithms that exhibit efficient multi-GPU
parallelism and low memory overhead, together with suitable domain-decomposition
and communication strategies. Such methods must also remain portable across
heterogeneous computing platforms and mainstream accelerator architectures \cite{leiserson2020there,trott2021kokkos,pennycook2019implications,hagerty2023studying,johansson2025lammpskokkos,gao2026rbmd}.
Addressing these closely coupled algorithmic and architectural requirements is
essential for advancing molecular dynamics into the exascale era \cite{pall2020heterogeneous,doijade2025redesigning,shaw2021anton3}.

In recent years, a class of stochastic methods has shown considerable promise for large-scale molecular dynamics simulations. A representative example is the random batch method \cite{jin2020random}, which has led to the development of the random batch Ewald (RBE) method \cite{jin2021random,liang2022superscalability,liang2022random} for long-range interactions and the random batch list (RBL) method \cite{liang2021random} for short-range interactions, with both subsequently applied to a variety of practical molecular systems \cite{liang2023random,chen2025random,gao2026symmetry,zhang2026random}. RBE employs importance sampling to avoid the communication-intensive FFT operations associated with mesh-based methods, thereby substantially improving the efficiency of large-scale parallel simulations on CPUs. RBL further introduces a core-shell decomposition of the near-field neighborhood, replacing the evaluation of a large number of particle interactions in the shell region by interactions with only a small randomly sampled subset, thus reducing both the computational cost and memory footprint of short-range force evaluation. These methods were subsequently accelerated on heterogeneous CPU-GPU architectures and incorporated into the RBMD simulation package \cite{gao2026rbmd}, resulting in further improvements in single-GPU performance and memory efficiency. A key remaining challenge is how to fully exploit the combined advantages of random batch methods and heterogeneous acceleration by integrating them with efficient multi-GPU parallelization strategies, thereby enabling simulations at substantially larger particle scales. Addressing this challenge is the central focus of the present work.

In this work, we develop RBMD 2.0, a major new release of the random batch molecular dynamics package \cite{gao2026rbmd}, designed for parallel execution across multiple GPU-accelerated nodes. It extends the previous single-GPU implementation to multi-node execution and supports systems with long-range interactions and complex force fields. A central feature of RBMD 2.0 is the co-design of random batch methods with multi-GPU communication and data-distribution strategies. The RBE implementation replaces FFT-based mesh redistribution by compact global reductions, whereas the RBL implementation combines three-dimensional domain decomposition with local ghost-particle exchange. At the software level, RBMD 2.0 replaces the VtK-m-based execution layer of the previous release \cite{gao2026rbmd,moreland2016vtkm} with the DTK CUDA architecture \cite{dtk2604cuda}, together with device-resident data structures and accelerator-aware communication interfaces. These developments extend the computational and memory advantages of random batch methods from single-device execution to multi-node GPU clusters. Numerical experiments on several complex systems with long-range interactions demonstrate  accurate reproduction of the target structural properties of all systems considered, while retaining linear scaling with system size. 
RBMD 2.0 exhibits nearly ideal weak scaling up to 16 nodes (64 DCUs). For the all-atom SPC/E water and PEO--LiTFSI systems, it achieves speedups of approximately $2.3$--$8.5\times$ over LAMMPS-Kokkos \cite{johansson2025lammpskokkos}, together with approximately two orders of magnitude acceleration for the long-range-dominated LJ electrolyte. These results demonstrate the attractive features of the new version of RBMD toward routine all-atom simulations of ultra-large charged soft-matter systems and molecular fluids at scales that are difficult to access efficiently using conventional multi-GPU molecular dynamics workflows.

\section{Method}
\label{sec::method}
\begin{figure*}[t] 
\centering   \includegraphics[width=1.0\textwidth]{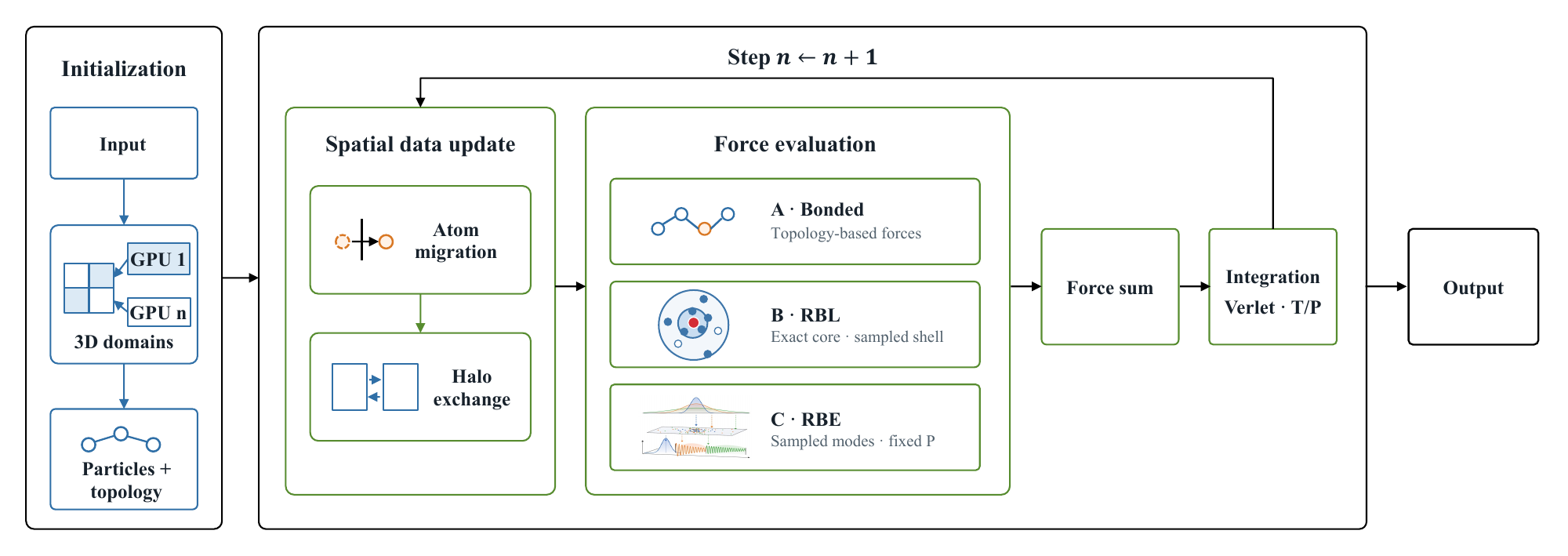}
\vspace{-0.2cm}
\caption{Schematic of the RBMD 2.0 workflow. The long-range and short-range components are evaluated with the random batch Ewald method and the random batch list method, respectively, adopting multi-GPU domain decomposition and initialization of the particle topology.}
\vspace{0cm}
\label{fig::RBMD_XS}
\end{figure*}

In this section, we introduce the simulation procedure for charged systems, the fast algorithms employed in RBMD 2.0, and their corresponding high-performance implementations. A schematic illustration for the workflow is provided in Figure~\ref{fig::RBMD_XS}.

\subsection{MD simulation and force evaluation}
\label{sec::model}

The molecular dynamics simulation workflow is well defined: after the particle information, force-field parameters, and simulation conditions are initialized, the forces acting on all particles are evaluated at each time step, and the system is subsequently advanced using a prescribed time integrator until the simulation is completed. The most computationally demanding and algorithmically complex component is the evaluation of particle interactions, whose mathematical form is determined by the selected force field \cite{Osguthorpe1998Structure,wang2004gaff,oostenbrink2004gromos,brooks1983charmm,jorgensen1996opls,sun1998compass}. For configuration $\bm{r}=(\bm{r}_1,\cdots,\bm{r}_N)$ and charge $\bm{q}=(q_1,\cdots,q_N)$, with charge neutrality condition $\sum_{i=1}^{N}q_i=0$, the corresponding potential energy is given by
\begin{equation}
   \label{eq::potential}
   U(\{\bm{r},\bm{q}\}   )=U_{\text{Bond}}+U_{\text{LJ}}+U_{\text{Coul}},
\end{equation}
where the three terms represent bonded, Lennard-Jones (LJ), and Coulomb interactions, respectively. The bonded potential $U_\text{Bond}$ arises from  bonds, bond angles, dihedral angles, as well as improper dihedral \cite{Osguthorpe1998Structure}. Although these interactions are local and involve only the immediate topological neighborhood of each particle, they play an essential role in preserving the molecular structure of the system. The LJ potential $U_{\text{LJ}
}$ reads
\begin{equation}
    \label{eq::LJ}
    U_{\text{LJ}
}=\sum_{i<j}^{N}4\epsilon_{ij}\left[\left(\frac{\sigma_{ij}}{r_{ij}}\right)^{12}-\left(\frac{\sigma_{ij}}{r_{ij}}\right)^{6}\right],
\end{equation}
with $r_{ij}=|\bm{r}_i-\bm{r}_j|$, where $\epsilon_{ij}>0$ and $\sigma_{ij}>0$ denote the characteristic energy and length scales, respectively. The LJ potential accounts for short-range repulsion and dispersion interactions. To mitigate finite-volume effects, molecular dynamics simulations commonly employ three-dimensional periodic boundary conditions \cite{frenkel2002understanding}. Under these conditions, the short-range LJ potential can be evaluated using the minimum-image convention, whereby, for each particle pair $(\bm{r}_i,\bm{r}_j)$, the interparticle separation $r_{ij}$ is taken as the closest distance among all periodic images. The Coulomb interaction term 
\begin{equation}
    \label{eq::Coul_potential}
    U_{\text{Coul}}=\frac{1}{2}\sum_{\bm{n}\in \mathbb{Z}^3}\sum_{i,j=1}^{N}\prime \frac{q_iq_j}{r_{ij}^{\bm{n}}},\  r_{ij}^{\bm{n}}:=|\bm{r}_i-\bm{r}_j+\bm{n}\circ\bm{L}|
\end{equation}
is indeed the most challenging part in MD simulation, where $\bm{L}=(L_x,L_y,L_z)$ denotes simulation-box dimension, `$\circ$' denotes the Hadamard product, and the prime indicates exclusion of the self-interaction term with $i=j$ and $\bm n=\bm 0$. The long-ranged nature and conditional convergence of the Coulomb interaction preclude the use of a finite cutoff, as is commonly employed for bonded and LJ interactions; instead, contributions from all particle pairs must be taken into account. The Ewald method provides a classical treatment of this problem by screening each point charge with a Gaussian charge distribution, thereby decomposing the Coulomb potential into the short- and long-range components \cite{ewald1921berechnung,frenkel2002understanding}, 
\begin{equation}
    \label{eq::Ewald}
    \frac{1}{r}=\frac{\erfc(\sqrt{\alpha}r)}{r}+\frac{\erf(\sqrt{\alpha}r)}{r},
\end{equation}
where $\erf(x)=2\int_{0}^x e^{-s^2}\mathrm{d}s/\sqrt{\pi}$ is the error function, $\erfc(x)=1-\erf(x)$ is the complementary error function, and $\alpha$ is a positive parameter that balances the two components. $U_{\text{Coul}}$ is thus decomposed into $U_{\text{Coul}}^{\mathcal{N}}+U_{\text{Coul}}^{\mathcal{F}}-U_{\text{Coul}}^{\mathcal{S}}$.
Owing to the rapid decay of the kernel $\operatorname{erfc}(\sqrt{\alpha}r)/r$, the short-range contribution $U_{\mathrm{Coul}}^N$ can be treated together with the LJ potential and evaluated using a finite cutoff. By contrast, the $\erf$ contribution $U_{\text{Coul}}^{\mathcal{F}}$ is long-range but sufficiently smooth to be handled efficiently in Fourier space, where the corresponding component takes the form
\begin{equation}
    \label{eq::U_Coul_Fourier}
    U_{\text{Coul}}^{\mathcal{F}}=\frac{2\pi}{V}\sum_{\bm{k}\neq \bm{0}}\frac{|\rho(\bm{k})|^2}{k^2}e^{-k^2/4\alpha},\  \rho(\bm{k}):=\sum_{j=1}^{N}q_je^{i\bm{k}\cdot \bm{r}_j},
\end{equation}
with box volume $V=L_xL_yL_z$ and the reciprocal space $\bm{k}\in \mathcal{K}:= 2\pi\mathbb{Z}^{3}\circ \bm{L}^{-1}$. The self-energy 
\begin{equation}
    \label{eq::U_self}
    U_{\text{Coul}}^{\mathcal{S}}=\sqrt{\frac{\alpha}{\pi}}\sum_{i=1}^{N}q_i^2
\end{equation}
corrects the self-interaction term in long-range Fourier calculation.
Eq.~\eqref{eq::U_Coul_Fourier} can be evaluated efficiently using FFT-based methods \cite{hockney1988computer,Darden1993JCP,essmann1995smooth}; however, within a parallel computing framework, such methods require substantial global communication among processing cores and across compute nodes, and this intrinsic algorithmic limitation constitutes a major obstacle to large-scale simulations. The challenge becomes even more pronounced in multi-GPU environments: a single GPU may support thousands of concurrent threads, while global communication among GPUs and across nodes in GPU clusters is considerably more difficult to implement efficiently.

To overcome the intrinsic limitations of Fourier mesh-based methods, RBMD 2.0 employs the improved random batch Ewald method \cite{liang2022improved} for evaluating interparticle interactions. This approach jointly accelerates the long- and short-range components: the long-range contribution is treated using the random batch Ewald (RBE) method \cite{jin2021random}, which replaces mesh-based FFTs with mesh-free importance sampling to reduce communication overhead and accelerate computation; the short-range contribution is handled using the random batch list (RBL) method \cite{liang2021random}, which further reduces the costs of neighbor-list construction and interaction evaluation.

At each simulation time step, the force $\bm{F}_i$ acting on the particle at position $\bm{r}_i$ is obtained as the negative gradient of the potential energy with respect to $\bm{r}_i$. Based on the Ewald decomposition, the nonbonded force is evaluated as $\bm{F}_i=\bm{F}_i^{\mathcal{N}}+\bm{F}_i^{\mathcal{F}}$, where
\begin{equation}
    \label{eq::F_i}
    \begin{aligned}
    \bm{F}_i^{\mathcal{N}}&=-\sum_{\bm{n},j}\prime\left(q_iq_jG(r_{ij}^{\bm{n}})+G_{\text{LJ}}^{ij}(r_{ij}^{\bm{n}})\right)\bm{r}_{ij}^{\bm{n}},\\
    \bm{F}_i^{\mathcal{F}}&=-\sum_{\bm{k}\neq \bm{0}}\frac{4\pi q_i\bm{k}}{Vk^2}e^{-k^2/4\alpha}\text{Im}\left(e^{-i\bm{k}\cdot \bm{r}_i}\rho(\bm{k})\right),
    \end{aligned}
\end{equation}
with
\begin{equation}
    \label{eq::G_short}
    \begin{aligned}
      &G(r):=\frac{\erfc(\sqrt{\alpha}r)}{r^3}+\frac{2\sqrt{\alpha}e^{-\alpha r^2}}{\sqrt{\pi}r^2},\\
      &G_{\text{LJ}}^{ij}(r):=24\epsilon_{ij}\left(\frac{2\sigma_{ij}^{12}}{r^{14}}-\frac{\sigma_{ij}^6}{r^8}\right).\\
    \end{aligned}
\end{equation}

For the long-range force $\bm{F}_i^{\mathcal{F}}$, the RBE method employs importance sampling to recast the sum over Fourier modes as the expectation of a random variable, which is then estimated using randomly sampled modes. The $P\in \mathbb{Z}_{+}$ i.i.d. random Fourier modes $\{\bm{k}_\ell\}_{\ell=1}^{P}$ are sampled from the discrete Gaussian distribution $\mathcal{P}(\bm{k})\sim e^{-k^2/4\alpha}$, $\bm{k}\in \mathcal{K}\backslash\{\bm{0}\}$, and the long-range force estimator reads
\begin{equation}
    \label{eq::F_long_RBE}
    \bm{F}_i^{\mathcal{F}}\approx\bm{F}_i^{\mathcal{F},*}:=\frac{S}{P}\sum_{\ell=1}^{P}\frac{4\pi q_i\bm{k}_\ell}{Vk_\ell^2}\text{Im}\left(e^{-i\bm{k}_\ell\cdot \bm{r}_i}\rho(\bm{k}_\ell)\right),
\end{equation}
where $S$ denotes the normalization constant of $\mathcal{P}(\bm{k})$. Previous studies have established that the required batch size $P$ is independent of the system size $N$ when its density $\rho_{\mathrm{sys}}$ is fixed \cite{jin2021random,liang2022superscalability,liang2023random,liang2022random,jiang2026random}, thus yielding an $O(N)$ complexity for the long-range calculation.

For the short-range force $\bm{F}_i^{\mathcal{N}}$, the neighbor list used in direct cutoff calculations may contain a large number of particles, often leading to memory-capacity and memory-access bottlenecks in heterogeneous multi-CPU and multi-GPU computing environments. The introduction of the RBL method can substantially reduce the number of neighboring particles required for the computation. Let
\begin{equation}
    \label{eq::Ci_Si}
    \mathcal{C}_i:=\{j\neq i|\ r_{ij}^{\bm{n}}\le r_c\},\ \mathcal{S}_i:=\{j\neq i|r_c< r_{ij}^{\bm{n}}\le r_s\}
\end{equation}
where $r_c$ and $r_s$ denote the user-defined core and shell cutoffs, respectively, under the minimum-image convention. Instead of evaluating all neighbors within the cutoff, RBL approximates the short-range force as
\begin{equation}
    \label{eq::RBL}
  \widetilde{\bm{F}}_{i}^{\mathcal{N},*}=\left[\sum_{j\in \mathcal{C}_i}+\frac{\#\mathcal{S}_i}{p}\sum_{j
  \in \mathcal{B}_i^{p}}\right]\bm{F}^{\mathcal{N}}_{ij},
\end{equation}
where $\bm{F}^{\mathcal{N}}_{ij}$ denotes the corresponding pair interaction in Eq.~\eqref{eq::F_i}, 
$\mathcal{B}_i^{p}\subset \mathcal{S}_i$ is a $p$-element subset sampled uniformly from $\mathcal{S}_i$, and `$\#$' denotes the cardinality of the particle set. A momentum-conserving correction then yields
\begin{equation}
    \label{eq::F_corr}
    \bm{F}_{i}^{\mathcal{N}}\approx\widetilde{\bm{F}}_{i}^{\mathcal{N},*}-\frac{1}{N}\sum_{j=1}^{N}\widetilde{\bm{F}}_{j}^{\mathcal{N},*}.
\end{equation}
This RBL estimator reduces the per-particle computational and memory costs from $O(4\pi\rho_{\rm sys}r_s^3/3)$ to $O(4\pi\rho_{\rm sys}r_c^3/3+p)$. In general, the variance decays rapidly with $r_c$, allowing $r_c$ to be chosen as approximately one-half of $r_s$ while maintaining sufficient accuracy \cite{liang2021random,liang2022improved,zhang2026random}, thereby yielding a substantial improvement in the efficiency of short-range force calculations.

RBMD 2.0 also provides sum-of-Gaussians (SOG) based random batch schemes for both short- and long-range interactions \cite{predescu2020u,liang2023random,jiang2026random}. In addition, the package supports multiple force-field models \cite{senftle2016reaxff,behler2007neural}, allowing users to select the desired force-evaluation scheme through the input file.

\subsection{DTK CUDA architecture and distributed data management}

\begin{figure*}[t] 
\centering   \includegraphics[width=1.0\textwidth]{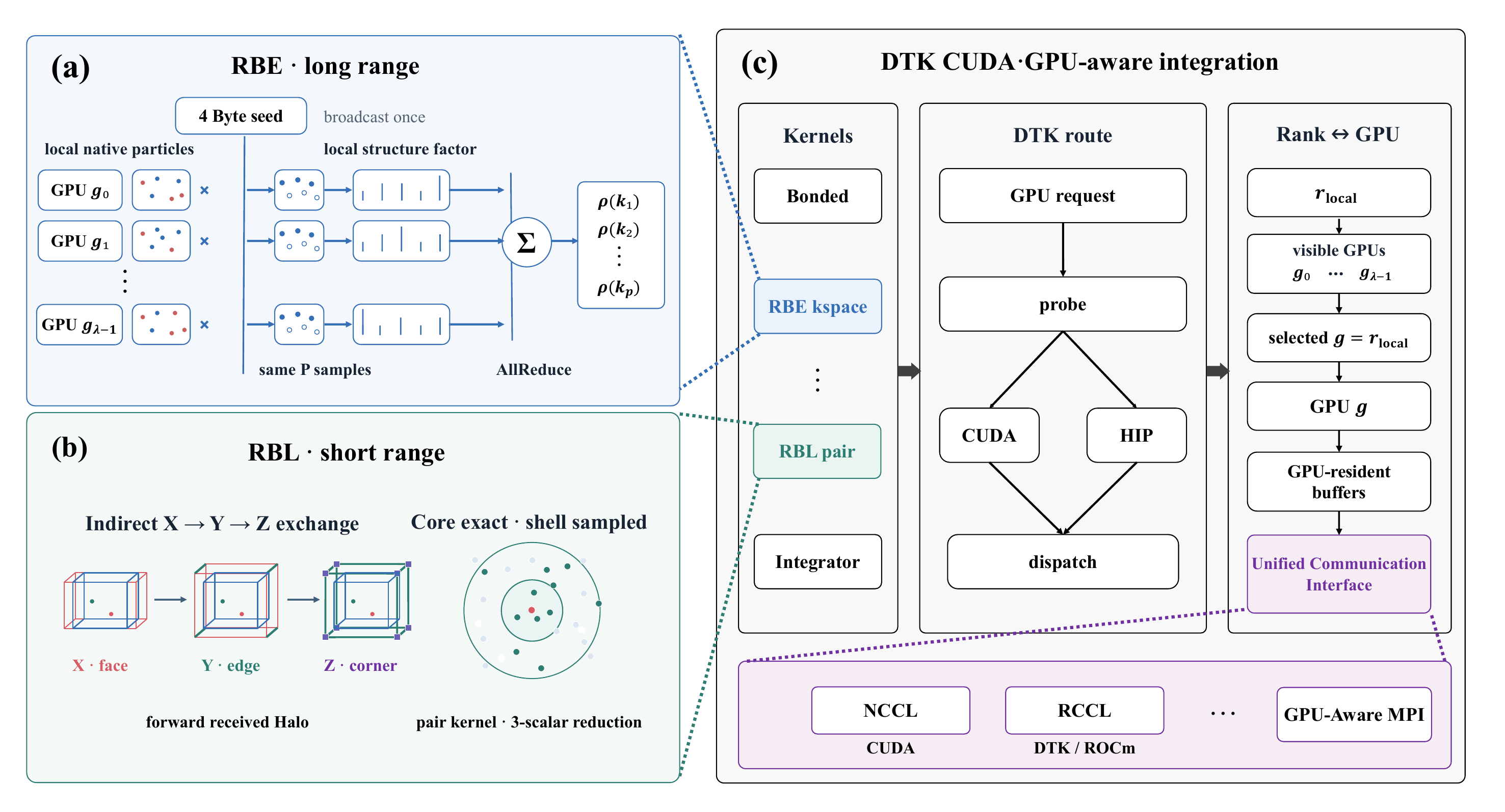}
\vspace{-0.2cm}
\caption{Overview of the multi-GPU implementation of RBMD 2.0. (a) RBE employs compact global structure-factor reductions, with communication volume independent of the system size. (b) RBL is integrated with domain decomposition, particle migration, and halo exchange to improve computational and memory efficiency. (c) DTK CUDA provides the common execution and GPU-aware communication layer for heterogeneous accelerator backends.}
\vspace{0cm}
\label{fig::multi_gpu}
\end{figure*}

To translate the algorithmic advantages of RBE and RBL into efficient large-scale simulations on heterogeneous accelerators, RBMD 2.0 adopts the DTK CUDA architecture \cite{dtk2604cuda} as the foundation of its high-performance implementation, as illustrated in Figure~\ref{fig::multi_gpu}(c). While retaining the data-parallel organization developed in the original RBMD 1.0 \cite{gao2026rbmd}, RBMD 2.0 replaces the VtK-m-based execution layer \cite{moreland2016vtkm} of the previous release with DTK CUDA. The resulting execution architecture separates high-level physical operators from device-specific kernel launch, memory management, and synchronization. RBMD adopts a backend-abstracted device programming model with native backends for DTK on DCU platforms, ROCm on AMD GPUs, and CUDA on NVIDIA GPUs, and the design is also extensible to additional accelerator backends. Consequently, RBE, RBL, bonded-force evaluation, integration, constraints, and particle reordering share a common high-level implementation over device arrays across DTK/DCU and CUDA-facing builds. This common implementation does not preclude architecture-aware optimization. For example, reductions and neighbor traversals follow DTK's native 64-lane execution semantics rather than assuming the 32-thread warp semantics of NVIDIA GPUs. Hardware-specific tuning can therefore be confined to a small set of kernels and adapters without duplicating the physical solvers, reducing the effort required to port, maintain, and validate the complete MD workflow on different accelerator platforms.

Particle attributes are stored in a structure-of-arrays (SoA) layout and remain device-resident together with linked-cell indices, improved random batch Ewald workspaces, force arrays, and communication buffers. Force evaluation and communication packing therefore operate directly on the same device-resident data, minimizing host-device transfers during the simulation. The distributed data path follows the same principle. Particle exchange is performed successively along the $x$, $y$, and $z$ directions, allowing edge and corner neighbors to be reached without explicit communication with every adjacent rank. For each transfer, a compact header first communicates the particle count and payload size, enabling the receiving rank to resize its buffer when necessary before transferring the contiguous SoA payload. Leaving particles are marked during the staged exchange and compacted only after halo construction, after which the particle arrays are reassigned and reordered according to linked-cell keys. Stable particle identifiers ensure that bonds, angles, dihedrals, impropers, and special-neighbor references can be migrated and reconstructed consistently with the particle data. These design choices combine portability with a GPU-efficient, device-resident data path, rather than treating multi-GPU support as a host-side wrapper around single-device kernels.

Above the execution layer, RBMD 2.0 provides typed \texttt{Broadcast}, \texttt{AllReduce}, \texttt{Reduce}, and \texttt{AllGather} operations through a vendor-independent collective interface. The current implementation supports MPI and NCCL-compatible backends, using NCCL on NVIDIA systems and the RCCL-compatible API on DTK/ROCm systems; the same abstraction is designed to accommodate oneCCL and HCCL backends without modifying the force solvers \cite{amdrccldocs,nvidianccldocs,oneccldocs,huaweihccldocs}. During initialization, each MPI rank is associated with a visible accelerator device, while GPU-Aware MPI is responsible for process bootstrap and geometrically local domain exchange. Eligible collective operations, by contrast, operate directly on device-resident buffers through accelerator-native communication backends. The accelerator-native backend uses an independent device stream, providing the capability for communication-computation overlap when permitted by the dependency graph. This execution and communication architecture provides the common infrastructure on which the algorithm-specific multi-GPU implementations of RBE and RBL are constructed, as described in Section~\ref{eq::imple_rbl_rbe}.

\subsection{Multi-GPU implementation of nonbonded interactions}
\label{eq::imple_rbl_rbe}

\subsubsection{Compact global structure-factor reduction}
The long-range component of Coulomb interactions is approximated by the RBE algorithm, which is implemented as a sampling, local accumulation, global reduction, and force-application pipeline, as shown in Figure~\ref{fig::multi_gpu}(a). Whenever the reciprocal mini-batch is refreshed, the root rank broadcasts one unsigned random seed, which occupies four bytes in the present implementation. Every rank then reconstructs the same ordered Fourier sample locally from the seed and global box parameters. The sampled vectors themselves are not communicated. Mode-wise sampling and particle-wise structure-factor evaluation are expressed through the DTK CUDA execution interface to exploit the available CPU or accelerator cores.

Each rank computes the real and imaginary structure-factor contributions using only its native particles, thereby avoiding duplicate contributions from ghosts. The many-core kernel accumulates these terms into two device-resident arrays, each containing $P$ real numbers. A device-aware \texttt{AllReduce} sums both arrays over all ranks, after which every accelerator applies the global structure factors to its native particles. The collective payload is therefore only two arrays of length $P$ and does not grow with the particle number or an FFT grid. RBE consequently avoids particle-to-mesh redistribution, mesh transposes, and distributed three-dimensional FFTs, which is particularly advantageous across nodes.

\subsubsection{Local halo exchange and neighbor construction}

The near-field nonbonded interactions comprise the short-range Coulomb contribution and the Lennard-Jones interactions. Their evaluation is coupled directly to the RBL neighborhood construction, as shown in Figure~\ref{fig::multi_gpu}(b). The global periodic box is divided into approximately balanced Cartesian subdomains \cite{PLIMPTON19951}. Each rank stores its native particles followed by a separate ghost range in the same structure-of-arrays containers; the halo thickness covers the short-range cutoff and neighbor-list skin. Native and ghost particles can therefore use one linked-cell traversal path.

Communication proceeds successively along the $x$, $y$, and $z$ directions. Each direction performs a leaving-particle exchange followed by a halo exchange, and particles received in one direction can be forwarded in the next. Edge and corner neighbors are thus covered without separate direct exchanges. Boundary tests and packing are performed on the accelerator using particle marks and a parallel scan. Each transfer sends a compact header containing the particle count and payload size before the SoA payload, allowing receive buffers to be reused or enlarged only when required.

Leaving particles carry the state, metadata, and bonded topology required to transfer ownership, whereas halo particles carry only the read-only fields required by the short-range kernels. Removal of leaving particles is deferred until all directional exchanges have finished. A single device-side compaction then updates the particle and topology arrays consistently, followed by linked-cell reassignment and cell-wise reordering. This avoids repeated reshuffling during one exchange cycle and restores contiguous access for force evaluation.

The RBL kernel constructs the core and shell neighborhoods of each native particle from the unified native-ghost cells. Ghost particles participate in both the exhaustive core calculation and stochastic shell sampling, preventing missing interactions at subdomain boundaries, but only native particles are integrated. After local short-range forces are obtained, an \texttt{AllReduce} of their three Cartesian sums supplies the same mean-force correction to every rank. RBL communication is therefore dominated by neighboring halo exchange, with only a three-scalar global correction. In combination, the implementation assigns local communication to RBL and a compact global collective to RBE, avoiding both complete shell neighbor lists and FFT mesh transposes.

\section{Results}
\label{sec::num_result}

We present a series of numerical examples for  RBMD 2.0 to demonstrate that it achieves substantial efficiency gains in multi-GPU parallel settings while maintaining accuracy. The benchmark suite includes long-range implicit-solvent systems, all-atom water systems for which structural preservation is more demanding, and more complex heterogeneous polymer systems. These test cases are described in detail in Section~\ref{sec::experiment}. For comparison, we use LAMMPS \cite{thompson2022lammps} (version 22Jul2025) with the Kokkos package \cite{trott2021kokkos} (version 4.6.2) to provide a fair multi-GPU parallel baseline. In LAMMPS-Kokkos, long-range electrostatics are evaluated using PPPM with an accuracy tolerance of $10^{-4}$. All simulations are performed in the canonical (NVT) ensemble using the Nos\'{e}–Hoover thermostat \cite{hoover1985canonical}. All benchmark tests are performed on a Hygon cluster. Each compute node contains two Hygon C86 processors (128 physical CPU cores in total, 2.5\,GHz, and 512\,GB of memory) and eight Hygon BW-series DCUs (gfx936 architecture and 64\,GiB HBM per device), which are designed for scientific computing rather than graphics rendering. In our experiments, 4 DCUs and 4 MPI ranks are used per node, with 1 MPI rank assigned to each DCU. All implementation of RBMD 2.0, as well as data for these numerical results, are available at \cite{RBMD2_0_release}.

\subsection{Experimental setup and metrics}
\label{sec::experiment}

The first system is implicit-solvent LJ electrolyte, which provides a simple yet representative model involving long-range electrostatic interactions and thus serves as a direct validation of the accuracy of both the algorithm and its implementation. All physical quantities are expressed in LJ reduced units. The system consists of $N=200,000$ monovalent ions in a cubic simulation box of side length $L=271$, with equal numbers of positively and negatively charged ions. The LJ parameters are set to $\epsilon_{ij}=\sigma_{ij}=1$ for all particle pairs, and the short-range cutoff is chosen as $r_s=8$. The temperature is maintained at $T=1$ using a thermostat, and the equations of motion are integrated with a time step of $\Delta t=0.001$.

The second system is all-atom SPC/E water, which involves both long-range electrostatic and bonded interactions and is therefore more sensitive to the accuracy of the force-evaluation algorithm in terms of structural properties. The system consists of $N=72,981$ atoms, corresponding to $24,327$ water molecules, in a cubic simulation box of side length $L=90~\mathrm{\AA}$. In the SPC/E model, the hydrogen atoms participate only in bonded and electrostatic interactions, whereas the LJ interaction acts only between oxygen atoms with parameters $\epsilon_{\mathrm{OO}}=0.1556~\mathrm{kcal/mol}$ and $\sigma_{\mathrm{OO}}=3.166~\mathrm{\AA}$ \cite{Berendsen2002missing}. The short-range cutoff is set as $r_s=9~\mathrm{\AA}$, the temperature is maintained at $T=298~\mathrm{K}$, and the equations of motion are integrated with a time step of $\Delta t=0.5~\mathrm{fs}$. The SHAKE algorithm \cite{krautler2001fast} is employed to constrain the intramolecular geometry of the water molecules.

The third system is PEO-LiTFSI polymer electrolyte, which is substantially more complex than the preceding two systems. The simulation box has dimensions $(L_x,L_y,L_z)=(239.585,221.524,226.684)~\mathrm{\AA}$ and contains 50 PEO chains, each consisting of $2,270$ EO units, together with $11,350$ LiTFSI molecules, resulting in a total of $N=976,220$ atoms. As a representative system described by the OPLS force field, PEO-LiTFSI involves considerably more complex LJ and bonded interaction parameters than the previous test cases. We adopt the same force-field parameterization as in \cite{doherty2017revisiting,jorgensen1996opls}. The short-range cutoff is selected as $r_s=12~\mathrm{\AA}$, the temperature is set at $T=343~\mathrm{K}$, and the equations of motion are integrated with a time step of $\Delta t=0.5~\mathrm{fs}$.

\begin{figure*}[t] 
\centering   \includegraphics[width=0.85\textwidth]{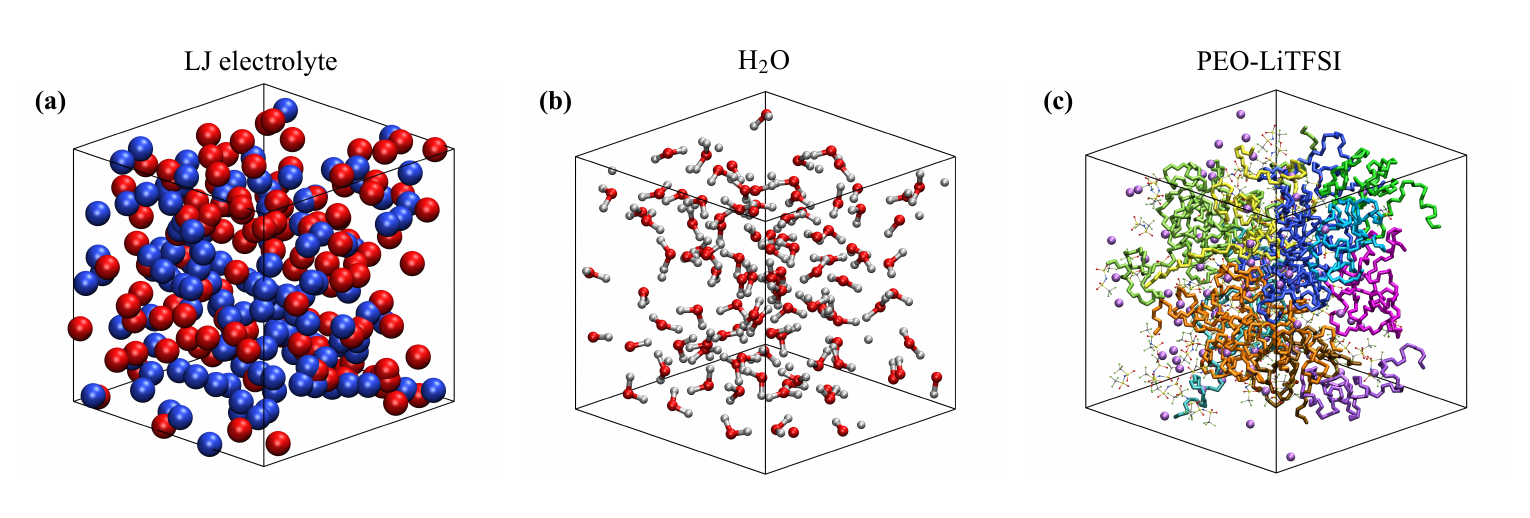}
\caption{System schematics of the three benchmark systems: (a) the LJ implicit-solvent electrolyte system, (b) the all-atom SPC/E water system, and (c) the PEO-LiTFSI polymer electrolyte system.}
\label{fig::system_sche}
\end{figure*}

Structural snapshots of these systems are shown in Figure~\ref{fig::system_sche}. Figure~\ref{fig::system_sche}(a) shows a snapshot of the LJ electrolyte, in which the cations and anions are represented by blue and red spheres, respectively, with no bonded interactions present. Figure~\ref{fig::system_sche}(b) shows the all-atom SPC/E water system, consisting of a large number of water molecules with explicit intramolecular bonds. Figure~\ref{fig::system_sche}(c) shows the PEO-LiTFSI polymer electrolyte, where long-chain PEO molecules serve as the solvent and are rendered in different colors for visual distinction. The $\mathrm{Li}^{+}$ ions are represented by purple spheres, while the bonded molecular structure of $\mathrm{TFSI}^{-}$ is explicitly displayed.

The experimental workflow is as follows. For each test system, once the system size, force field, topology, and short-range cutoff $r_s$ are fixed, we first use LAMMPS-Kokkos to compute structural observables, such as the radial distribution function (RDF), which serve as reference quantities for accuracy assessment. We then retain the same Verlet-list algorithm for short-range interactions in RBMD 2.0, replace the long-range solver by RBE, and progressively increase the batch size $P$ until the resulting RDF is in close agreement with the reference. Next, the short-range solver is replaced by RBL, and the core cutoff $r_c$ and batch size $p$ are adjusted to maintain the same level of agreement. This procedure determines the complete set of force-evaluation parameters $(r_s,r_c,p,P)$. The resulting force-evaluation parameters are summarized in Table~\ref{tab::rbmd_parameter}, and the same parameter set is subsequently used in all timing and scaling experiments to ensure both accuracy and fairness of the performance comparison. For the performance comparison, we measure the nonbonded force-evaluation time, including both the short-range and long-range components, using the average of $1,000$ evolution time steps. For both implementations, fully synchronized timing is employed to ensure consistent performance measurements across MPI ranks. LAMMPS-Kokkos is executed with the \texttt{timer full sync} option \cite{lammps_timer}, and an equivalent synchronization strategy is used in RBMD 2.0.

\begin{table}[htbp]
\caption{Force-evaluation parameters used for each benchmark system in RBMD 2.0. The LJ electrolyte uses LJ reduced units, while the other systems use real units.}
\label{tab::rbmd_parameter}
\renewcommand{\arraystretch}{1.3} 
\begin{ruledtabular} 
\begin{tabular*}{\columnwidth}{@{\extracolsep{\fill}}ccccc@{}}
System & $r_s$ & $r_c$ & $p$ & $P$\\ \hline
LJ electrolyte    & 8 & 4 & 30 & 200  \\
SPC/E water   & $9~\mathrm{\AA}$ & $5.5~\mathrm{\AA}$ & 50 & 200  \\
PEO-LiTFSI    & $12~\mathrm{\AA}$ & $6~\mathrm{\AA}$ & 200 & 200  \\
\end{tabular*}
\end{ruledtabular}
\end{table}

\subsection{Accuracy validation}
\label{sec::performace}

\begin{figure*}[t] 
\centering   \includegraphics[width=1.0\textwidth]{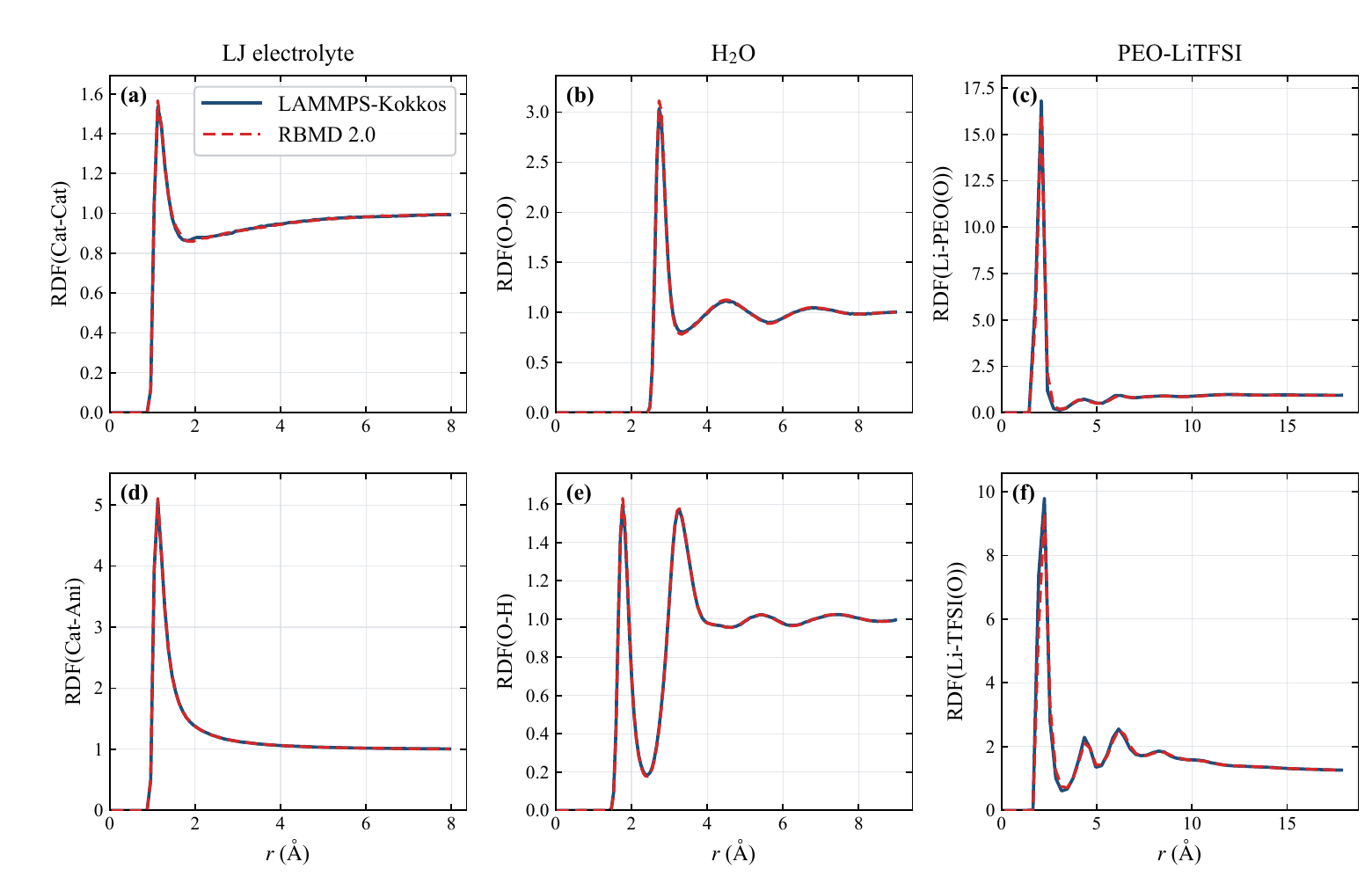}
\caption{RDF comparisons for the three benchmark systems: (a,d) the LJ implicit-solvent electrolyte system, (b,e) the all-atom SPC/E water system, and (c,f) the PEO-LiTFSI polymer electrolyte system. The parameters used in RBMD 2.0 are listed in Table~\ref{tab::rbmd_parameter}. LAMMPS-Kokkos uses the same short-range cutoff $r_s$, while long-range electrostatics are evaluated using PPPM with an accuracy tolerance of $10^{-4}$. All experiments are performed on 4 DCUs.}
\label{fig::rdf}
\end{figure*}

We first assess the accuracy of RBMD 2.0 through structural properties of the simulated systems. Here, we focus on the RDFs, which characterize the spatial distribution and local structural correlations of particles and are sensitive to errors in the force evaluation. For each benchmark system, the simulation is first equilibrated and subsequently evolved for an additional $2.0\times10^{4}$ time steps, over which the RDF is computed. The results obtained with the two software packages are presented in Figure~\ref{fig::rdf}, where two representative pairwise RDFs are reported for each benchmark system. For the LJ implicit-solvent electrolyte, Figures~\ref{fig::rdf}(a,d) show the cation-cation and cation-anion RDFs, respectively. For the SPC/E water system, Figures~\ref{fig::rdf}(b,e) show the O-O and O-H RDFs. For the PEO-LiTFSI system, Figures~\ref{fig::rdf}(c,f) show the distributions between Li ions and O atoms in PEO and LiTFSI, respectively. For RBMD 2.0, all RDFs show close agreement with the LAMMPS-Kokkos results. Since the accuracy and convergence of the underlying improved random batch method have been systematically established in previous studies \cite{jin2021random,liang2021random,liang2022improved,liang2022superscalability,gao2026rbmd}, the close agreement here confirms that the structural accuracy is preserved in the multi-GPU implementation of RBMD 2.0.

\begin{figure*}[t] 
\centering   \includegraphics[width=1.0\textwidth]{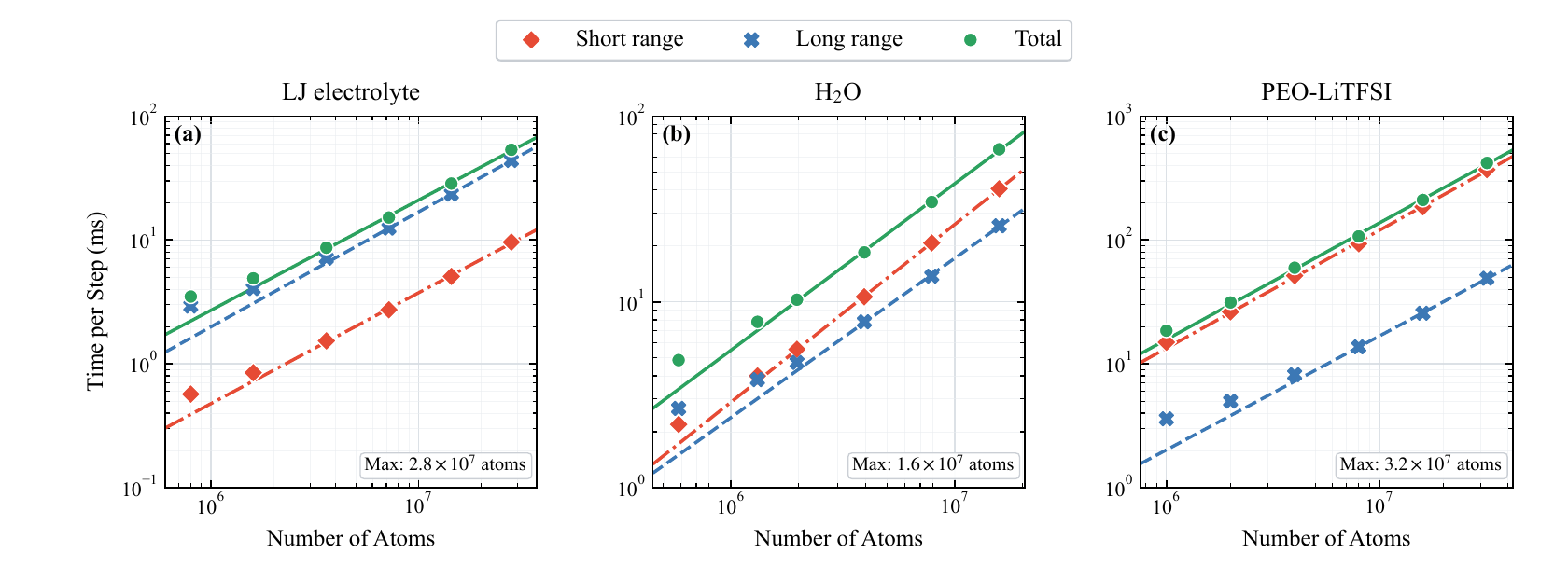}
\caption{Complexity verification of RBMD 2.0 for the benchmark systems. All results are obtained on a fixed single node with 4 DCUs. The solid markers denote the measured data, while the lines represent the fitting line of linear growth. 
}
\label{fig::linear_scaling}
\end{figure*}

\subsection{Performance and  parallel efficiency}
Next, we present a series of timing experiments to assess the computational efficiency of RBMD 2.0. We first examine whether the expected linear complexity is preserved in a multi-GPU parallel setting. The complexity results are shown in Figure~\ref{fig::linear_scaling}, where the system size spans more than one order of magnitude for all three benchmark systems, while the number of DCUs is fixed at four. Despite the distributed execution across multiple DCUs, RBMD 2.0 consistently exhibits $O(N)$ scaling for all systems, thereby enabling simulations at substantially larger scales.

\begin{figure*}[t] 
\centering   \includegraphics[width=0.6\textwidth]{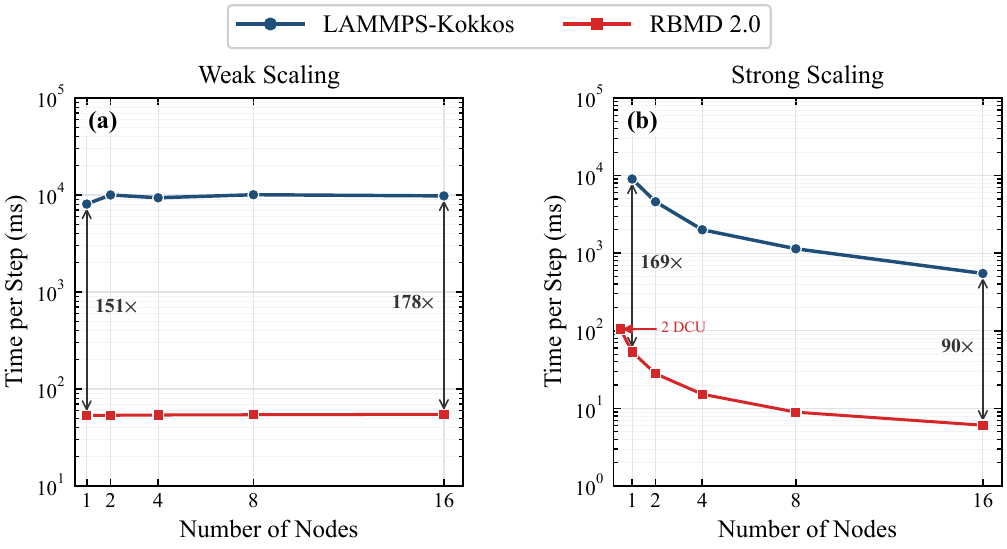}
\caption{Comparison of (a) weak scalability and (b) strong scalability for the LJ implicit-solvent electrolyte system using different software implementations. In the weak-scaling test, the number of atoms per DCU is fixed at $N_{\mathrm{DCU}}=7,000,000$, whereas in the strong-scaling test, the total system size is fixed at $N=28,000,000$. The RBMD 2.0 parameters are kept identical to those listed in Table~\ref{tab::rbmd_parameter}.}
\label{fig::lj_strong_weak}
\end{figure*}

We then examine the strong and weak scalability of RBMD 2.0, which is critical for assessing its capability to perform large-scale simulations across multiple accelerator devices. As the number of parallel devices increases, inter-GPU communication overhead can gradually become dominant over the computational cost, and this issue is generally more pronounced than in conventional CPU-based parallelization. Strong and weak scalability provide standard measures for quantifying this parallel efficiency. For weak scaling, the number of particles assigned to each device is kept fixed while both the number of devices and the total system size are increased. The weak-scaling efficiency is defined by  $\eta_{\text{weak}}(\lambda)=t(\lambda_{\text{ini}})/t(\lambda)$
where $t(\lambda)$ denotes the wall-clock time measured using $\lambda$ accelerator devices (or nodes), and $\lambda_{\mathrm{ini}}$ denotes the baseline configuration.
Ideally, the nonbonded-force-evaluation time should remain approximately constant as the number of devices increases.
Strong scaling is assessed by fixing the total system size while increasing the number of accelerator devices, and the strong-scaling efficiency is defined as 
$\eta_{\text{strong}}(\lambda)=t(\lambda_{\text{ini}})/t(\lambda)\cdot\lambda_{\text{ini}}/\lambda.$
 Ideally, the nonbonded-force-evaluation time decreases proportionally with the number of devices.
For large-scale simulations, it is important that each accelerator maintains a sufficiently large workload while parallel efficiency remains high as additional devices are introduced. Weak scaling is therefore particularly relevant in the present setting and serves as a primary measure of the scalability of RBMD 2.0. Similar scalability metrics have also been adopted in previous studies \cite{duan2020cell}.

The LJ implicit-solvent electrolyte contains only cations and anions and involves no bonded interactions. Consequently, its computational cost is dominated by the evaluation of long-range electrostatic interactions, which is substantially more expensive than the short-range component. The strong- and weak-scaling results for the two software packages are compared in Figure~\ref{fig::lj_strong_weak}, using one node with 4 DCUs as the baseline. Compared with LAMMPS-Kokkos, the mesh-free RBE implementation reduces the nonbonded force-evaluation cost by approximately two orders of magnitude. Figure~\ref{fig::lj_strong_weak}(a) further shows that RBMD 2.0 maintains nearly ideal weak-scaling behavior even at the largest tested system size of $N=2.56\times10^{8}$ particles, with its performance advantage over LAMMPS-Kokkos becoming more pronounced as the number of DCUs increases. For the strong-scaling test in Figure~\ref{fig::lj_strong_weak}(b), we additionally report the RBMD 2.0 result obtained using two DCUs for a fixed system size of $N=28,000,000$. Under the same configuration, the memory requirement of LAMMPS-Kokkos exceeds the available device memory, preventing the simulation from being completed. This comparison highlights the reduced memory footprint of the random-batch force-evaluation framework in RBMD 2.0, allowing substantially more particles to be accommodated on each accelerator device.

\begin{figure*}[t] 
\centering   \includegraphics[width=1.0\textwidth]{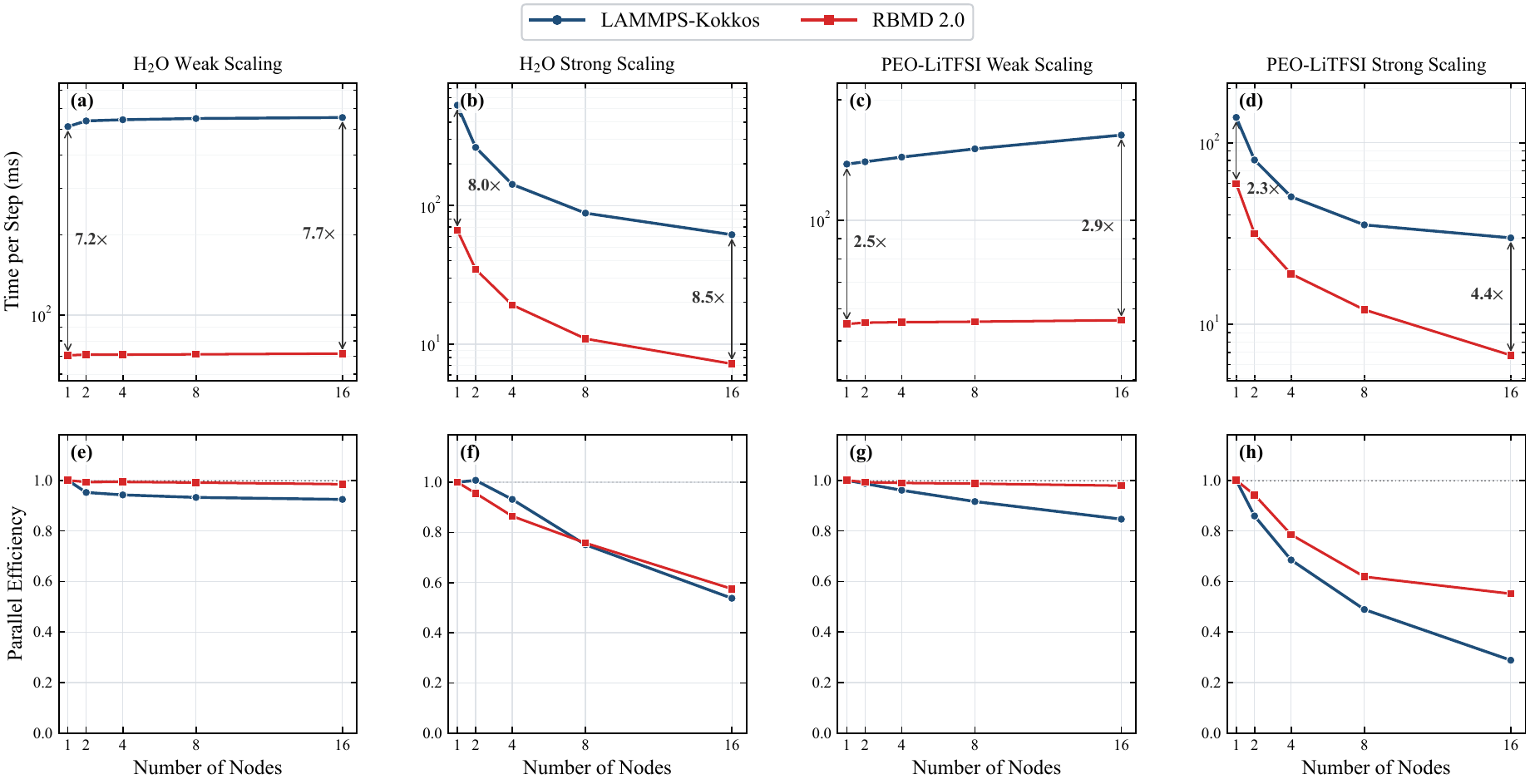}
\caption{Comparison of weak and strong scalability for (a,b,e,f) SPC/E all-atom water system and (c,d,g,h) PEO-LiTFSI system using different software implementations. In the weak-scaling test, the number of atoms per DCU is fixed at (a,e) $N_{\mathrm{DCU}}=3,940,974$ and (c,g) $N_{\mathrm{DCU}}=1,000,000$, whereas in the strong-scaling test, the total system size is fixed at (b,f) $N=15,763,896$ and (d,h) $N=4,000,000$, respectively. The RBMD 2.0 parameters are kept identical to those listed in Table~\ref{tab::rbmd_parameter}.}
\label{fig::h2o_peo_strong_weak}
\end{figure*}

For the all-atom SPC/E water and PEO-LiTFSI systems, short-range interactions become substantially more important, and their computational cost is often comparable to or even larger than that of the long-range component. Figure~\ref{fig::h2o_peo_strong_weak} presents the strong- and weak-scaling results for the two software packages, again using one node with four DCUs as the baseline. The results show that, even for these more complex systems where short-range force evaluation accounts for a larger fraction of the total cost, RBMD 2.0 achieves a speedup of approximately $2.3\,$-$\,8.5\times$ over LAMMPS-Kokkos. Figures~\ref{fig::h2o_peo_strong_weak}(a,c,e,g) demonstrate nearly ideal weak-scaling behavior with $\eta_{\text{weak}}\ge 97.5\%$ up to 16 nodes for RBMD 2.0, whereas Figures~\ref{fig::h2o_peo_strong_weak}(b,d,f,h) show that its performance advantage becomes increasingly pronounced in the strong-scaling regime. The reduced strong-scaling efficiency for RBMD 2.0 in Figure~\ref{fig::h2o_peo_strong_weak}(f,h) arises because the per-step force-evaluation time has already reached the millisecond scale, making the constant $O(P)$ structure-factor communication overhead increasingly significant. The largest simulations also reach system sizes on the order of $10^8$ particles. These results demonstrate that, while maintaining accuracy, RBMD 2.0 provides efficient multi-GPU parallel nonbonded force evaluation and suggest favorable scalability toward substantially larger, potentially exascale, molecular dynamics simulations.

\section{Conclusion}
\label{sec::conclusion}

In this work, we developed RBMD 2.0, a major new release of the RBMD molecular dynamics package for efficient large-scale simulations on multi-GPU clusters. By co-designing random batch force-evaluation algorithms with multi-GPU communication and data distribution, RBMD 2.0 reduces long-range communication to compact global collectives while confining short-range communication primarily to local halo exchanges. Numerical results for the LJ electrolyte, SPC/E water, and PEO-LiTFSI systems reproduce the structural properties obtained with LAMMPS-Kokkos and demonstrate nearly ideal weak scaling up to 16 nodes (64 DCUs). Compared with LAMMPS-Kokkos, RBMD 2.0 achieves severalfold speedups in nonbonded force evaluation for the all-atom systems and approximately two orders of magnitude acceleration for the long-range-dominated LJ electrolyte, with simulations reaching $2.56\times10^8$ particles. For complex all-atom systems, short-range interactions and neighbor processing increasingly dominate the computational cost and therefore limit further acceleration. Future work will focus on optimizing neighbor construction and traversal and extending the framework to larger heterogeneous accelerator platforms.

\section*{Acknowledgement}

This work is supported by the National Natural Science Foundation of China (Grants No. 12426304, 12325113 and 125B2023). The authors would like to thank the support from the SJTU Kunpeng \& Ascend Center of Excellence.

\section*{Conflict of interest}

The authors declare that they have no conflict of interest.

\section*{Data Availability Statement}

The source code of RBMD 2.0 and the data supporting the findings of this study
are openly available at
\url{https://github.com/SOG-AITech/rbmd}.


\end{document}